\documentclass[12pt,a4paper]{revtex4}

\usepackage{latexsym,amssymb,amsmath,amscd,amscd,amsthm,amsxtra,graphicx,psfig}
\usepackage[mathscr]{eucal}
\usepackage{mathrsfs}

\newcommand{\ba}{\begin{eqnarray}} \newcommand{\ea}{\end{eqnarray}}
\newcommand{\bc}{\begin{center}}   \newcommand{\ec}{\end{center}}

\begin{document}

%\preprint{}

\title{Twist-3 distribution amplitudes of the pion and kaon from the QCD sum rules }
\author{Tao Huang$^{1,2}$} \email{huangtao@mail.ihep.ac.cn}
\author{ Ming-Zhen Zhou$^2$} \email{zhoumz@mail.ihep.ac.cn}
\author{ Xing-Hua Wu$^2$} \email{xhwu@mail.ihep.ac.cn}
\affiliation{$^1$CCAST(World Laboratory), P.O.Box 8730, Beijing 100080, P.R.China}
\affiliation{$^2$Institute of High Energy Physics, P.O.Box 918, Beijing 100039, P.R.China}
\date{January 6, 2005}

\begin{abstract}
Twist-3 distribution amplitudes of the pion and kaon are studied in this paper.
We calculate the fist several moments for the twist-3 distribution amplitudes
($ \phi_{p,\sigma}^\pi$ and $ \phi_{p,\sigma}^K$)
of the pion and kaon by applying the QCD sum rules. Our results show that,
(i) the first three moments of $ \phi_p^K $ and the first two
moments of $ \phi_p^\pi $ and $ \phi_\sigma^{\pi,K} $ of the pion and kaon can be
obtained with 30$\%$ uncertainty;
(ii) the fourth moment of the $ \phi_p^\pi $ and the second moment of
the $ \phi_\sigma^K $ can be obtained when the uncertainty are relaxed to 35$\%$;
(iii) the fourth moment of the  $ \phi_\sigma^\pi $ can be obtained only
when the uncertainty are relaxed to 40$\%$;
(iv) we have $m_{0\pi}^p=1.10\pm 0.08~{\rm GeV}$ and
$m_{0K}^p=1.25\pm 0.15~{\rm GeV}$ after including the $\alpha_s$-corrections to the perturbative
part. These moments will be helpful for constructing the twist-3 wave functions of the pion and kaon.
\\
PACS number(s): 13.20.He 11.55.Hx
\end{abstract}

\maketitle

\newpage

\section{Introduction}

Hadronic distribution amplitudes, which involve the non-perturbative information,
are the important ingredients when applying the QCD to hard exclusive
processes via the factorization theorem. These distribution amplitudes are
process-independent and should be determined by the hadronic dynamics. They satisfy the
renormalization group equation and have the asymptotic solutions as $Q^2\to \infty$.

From the counting rule, the twist-2 distribution amplitude makes the leading contribution and the
contribution from the higher-twist distribution amplitude is suppressed by a
factor $1/Q^2$ in the large momentum
transfer regions. However as one wants to explain the present experimental data, the
non-leading contributions should be taken into account. The non-leading contributions include
higher-order corrections, higher-twist and higher Fork state contributions et.al. Therefore one
has to study the twist-2 and higher-twist distribution amplitudes as the universal nonperturbative
inputs for the exclusive processes.

Distribution amplitudes can be obtained from the hadronic wave functions by
integrating the transverse momenta of the quarks in the hadrons.
For example, the pionic distribution amplitudes of the lowest Fork state are defined as  :
\ba
&&\left< 0 \left| \bar d_\alpha(z)\left[z,-z\right] u_\beta (-z)\right|\pi(q) \right>
\nonumber \\
&&\quad
=-\,{i\over 8}f_\pi\int_{-1}^1d\xi ~ e^{i \xi(z \cdot q)}
  \left\{
         \not\! q \gamma_5 \phi_\pi(\xi)
         + m_{0\pi}^p \gamma_5\phi_p^\pi(\xi)
         + {2\over 3}m_{0\pi}^\sigma\sigma_{\mu\nu}\gamma_5
             q^\mu z^\nu \phi_\sigma^\pi(\xi)
  \right\}_{\beta\alpha}
+ \dots
%\nonumber\\
\label{pi-distribution-amplitudes}
\ea
where $ \sigma_{\mu\nu} = {i \over 2}(\gamma_\mu \gamma_\nu -\gamma_\nu \gamma_\mu) $,
$f_\pi$ is the pion decay constant and
\ba
\left[ z,-z \right]= \exp \left\{i g \int^{z}_{-z}dx^{\mu}A_{\mu} \right \}
\nonumber
\ea
is the Wilson line inserted to preserve gauge invariance of the distribution
amplitudes. The $\phi_\pi(\xi)$ ,
$\phi_p^\pi(\xi)$ and $\phi_\sigma^\pi(\xi)$ in Eq.(\ref{pi-distribution-amplitudes})
are the twist-2 and two twist-3 (non-leading) distribution amplitudes respectively.
For the $K$ meson, the definition is similar except for the $d$ quark replaced by
the $s$ quark and $m_{0\pi}^{p,\sigma}$ replaced by $m_{0 K}^{p,\sigma}$~.

To isolate the light-cone twist-3 distribution amplitudes $ \phi_p^\pi $ and
$ \phi_\sigma^\pi $ of the pion, one can
contract the equation (\ref{pi-distribution-amplitudes}) with the gamma
matrices $\gamma_5$ and $\sigma_{\mu\nu}\gamma_5$ respectively,
\ba
\left<  0\left|\bar{d}(z) i
\gamma_5 \left[ z,-z \right] u(-z)\right| \pi^+(q)\right> =m_{0\pi}^p f_\pi~
\frac{1}{2}\int_{-1}^{1}d\xi~ \phi_p^\pi(\xi) e^{i \xi(z \cdot q)}
+\cdots
\label{1}
\ea
and
\ba
\left<  0\left|\bar{d}(z) \sigma_{\mu\nu} \gamma_5 \left[ z,-z \right] u(-z)\right| \pi^+(q)\right>
&\makebox[0pt]{=}&{-i~ m_{0\pi}^\sigma f_\pi \over 3}(q_\mu z_\nu - q_\nu z_\mu)~
\frac{1}{2}\int_{-1}^{1}d\xi~ \phi_\sigma^\pi(\xi) e^{i \xi(z \cdot q)}
\nonumber \\
&\makebox[0pt]{}&+\cdots
\label{8}~ .
\ea
In a similar way, we can define two twist-3 distribution
amplitudes $ \phi_p^K $ and $ \phi_\sigma^K $ of the kaon in the following,
\ba
\left<  0\left|\bar s(z) i \gamma_5 \left[z,-z \right]u(-z)\right| K^+(q)\right>
= m_{0 K}^p f_K~\frac{1}{2}\int_{-1}^{1}d\zeta~ \phi_p^K(\zeta)
e^{i \zeta(z\cdot q)} +\cdots
\label{19}
\ea
and
\ba
\left< 0\left|\bar s(z)\sigma_{\mu \nu} \gamma_5  \left[ z,-z \right] u(-z)\right|K^+(q)\right>
&\makebox[0pt]{=}&{-i~ m_{0 K}^\sigma f_K \over 3}(q_\mu z_\nu - q_\nu z_\mu)~
\frac{1}{2}\int_{-1}^{1}d\zeta~\phi_\sigma^K(\zeta) e^{i \zeta(z \cdot q)}
\nonumber \\
&\makebox[0pt]{}&+ \cdots
\label{20} ~ .
\ea
The dots in the above definitions are refer to those higher-twist distribution amplitudes.
We do not consider their influences in the following calculation.

As pioneered in Ref.\cite{moment-qcd-sum-rule,collinear-exclusive-2}, the authors pointed
out that the first several moments of the distribution amplitudes could be calculated in the QCD
sum rules \cite{qcd-sum-rule}.
Those moments are helpful to construct a model for the hadronic distribution amplitudes.

The parameters $m_{0\pi}^{p,\sigma}$ and $m_{0 K}^{p,\sigma}$ introduced
in the definition are used to normalize the zeroth moments
of their corresponding distribution amplitudes.
It is shown in this paper that these parameters
determined by the QCD sum rules are smaller than those
required by the equations of motion
(e.g., see Ref. \cite{light-cone-non-leading-twist-def-1,light-cone-non-leading-twist-def-2}).

In this paper, we calculate the first three moments of twist-3 distribution amplitudes of
the $ \pi $ and $ K $ , defined in Eqs.(\ref{1})-(\ref{20}), in the QCD sum rules.
For the pion case, we had calculated the moments of distribution amplitude $ \phi_p^\pi $ in
the previous paper \cite{replace}. However there were some mistakes in estimating the contribution
from the continuous spectrum and the Borel windows which
would severely influence the values of the moments.
Now we present the correct expressions for the moments
of $\phi_p^\pi$ and re-analyse their numerical results in this paper.
Furthermore, it is well known that axial
 currents in a correlator would couple to
 instantons(see, for example, Ref.\cite{Baulieu}). And this may cause
 some complication in the calculation and make the results unreliable.
 We will not explore their influences in this paper.

This paper is organized as follows.
In Sec.II, we give the sum rules of the moments of $ \phi_p^\pi $ and $ \phi_\sigma^\pi $
for the $\pi$ meson.
The sum rules for the moments of $ \phi_p^K $ and $ \phi_\sigma^K $
of $ K $ meson are given in Sec.III.
The SU(3) symmetry violation have be taken into account.
In Sec.IV, numerical analysis of various moments is presented.
The information about the 3-particle twist-3 distribution amplitude obtained from
the 2-particle distribution amplitudes are also discussed.
The last section is reserved for summary and discussion.

\section{QCD sum rules for the moments of $ ~\phi_p^\pi~$ and $~\phi_\sigma^\pi~ $ of the pion}

In this section we apply the background field method in QCD
to calculate the moments
\cite{back-ground-field-1,back-ground-field-2,back-ground-field-3,back-ground-field-4,back-ground-field-5,back-ground-field-6}.
Expanding equations (\ref{1}) and (\ref{8}) around $ z^2=0 $, we have
\ba
\left< 0 \left| \bar{d}(0)\gamma_5(iz\cdot \overleftrightarrow{D})^{n} u(0) \right| \pi^+(q) \right>
= -i f_\pi m_{0\pi}^p \left< \xi_p^{n} \right> (z\cdot q)^{n}
\label{add5}
\ea
and
\ba
\left< 0 \left| \bar{d}(0)\sigma_{\mu \nu}\gamma_5
(iz\cdot \overleftrightarrow{D})^{n+1} u(0) \right| \pi^+(q) \right>
= -{n+1 \over 3}f_\pi m_{0\pi}^\sigma \left< \xi_\sigma^{n} \right>
(q_\mu z_\nu - q_\nu z_\mu) (z\cdot q)^{n}
\label{add6}
\ea
respectively. The moments in the equations (\ref{add5}) and (\ref{add6}) are defined by
the following expressions,
\ba
\left< \xi_p^{n}\right>=\frac{1}{2}\int_{-1}^{1}\xi^{n}\phi_p^\pi(\xi)d\xi~,
\quad\quad
\left< \xi_\sigma^{n}\right> = \frac{1}{2}\int_{-1}^{1}\xi^{n}\phi_\sigma^\pi(\xi)d\xi
\label{add9}~.
\ea
As usual, the $ SU(2) $ isospin
symmetry can be taken as (nearly) exact. It means that the
distribution of longitudinal momentum carried by the quarks (in the light cone framework)
should be symmetric between $ u $ and $ d $, i.e., odd moments of the distribution
amplitudes $ \phi_p^\pi, \phi_\sigma^\pi $ should be zero. So we consider only the even moments
for the pion case in the following.

In order to obtain the sum rules of the moments, we introduce two corresponding correlation functions,
\ba
(z\cdot q)^{2n}~I_p^{(2n,0)}(q^2) \equiv  -i\int d^4 x e^{iq\cdot x}
\left< 0 \left| {\rm T} \left \{ \bar{d}(x)\gamma_5(iz\cdot \overleftrightarrow{D})^{2n} u(x),
\bar u(0)\gamma_5 d(0) \right\} \right| 0 \right>
\label{10}
\ea
and
\ba
&& -i (q_\mu z_\nu - q_\nu z_\mu) (z\cdot q)^{2n}~I_\sigma^{(2n,0)}(q^2)
\nonumber \\
&&\quad \equiv -i\int d^4 x e^{iq\cdot x} \left< 0 \left| {\rm T}
\left \{ \bar{d}(x)\sigma_{\mu \nu}\gamma_5(iz\cdot \overleftrightarrow{D})^{2n+1} u(x),
\bar u(0)\gamma_5 d(0)\right\} \right| 0 \right>
\label{11}~.
\ea
In the deep Euclidean region($~ -q^2 \gg 0~ $), one can calculate the
Wilson coefficients in the operator product expansion
(OPE) for Eqs.(\ref{10}) and (\ref{11}) perturbatively.
The results with power correction to dimension six and the $\alpha_s$ corrections
to lowest order are written as
\ba
I_p^{(2n,0)}(q^2)_{\rm QCD} &\makebox[0pt]{=}&-{1
\over 2n+1}~{3 \over 8 \pi^2}q^2 \ln{-q^2 \over \mu^2} -{1 \over
8}{\left< \dfrac{\alpha_s}{\pi}G^2 \right> \over q^2} -{2n - 1
\over 2}{(m_u+m_d)\left< \bar q q\right> \over q^2}
\nonumber \\
&\makebox[0pt]{}& +{16\pi \over 81}(16n^2+4n+21){\left<
\sqrt\alpha_s \bar q q\right>^2 \over q^4}
\label{pi-p-qcd}
\ea
and
\ba
I_\sigma^{(2n,0)}(q^2)_{\rm QCD}
&\makebox[0pt]{=}& - {1 \over 2n+3}~{3 \over 8 \pi^2}q^2 \ln{-q^2 \over \mu^2}
 -{1 \over 24}{\left< \dfrac{\alpha_s}{\pi}G^2 \right> \over q^2}
 -{2n + 1 \over 2}{(m_u+m_d)\left< \bar q q\right> \over q^2}
\nonumber \\
&\makebox[0pt]{}& +{16\pi \over 81}(16n^2+12n-7)
                   {\left< \sqrt\alpha_s \bar q q\right>^2 \over q^4} .
\label{pi-sigma-qcd}
\ea
On the other hand, in the physical region, correlation functions (\ref{10})
and (\ref{11}) can be written in terms of their hadronic spectrum representation
(according to Eqs.(\ref{add5}) and (\ref{pi-p-qcd}),
(\ref{add6}) and (\ref{pi-sigma-qcd}) respectively),
\ba
{\rm Im} ~ I_p^{(2n,0)}(q^2)_{\rm had} =
\pi \delta(q^2-m_\pi^2)f_\pi^2 (m_{0\pi}^p)^2 \left< \xi_p^{2n}\right>
+\pi {3 \over 8 \pi^2}{1\over 2n+1}q^2\theta(q^2-s_\pi^p)
\label{pi-p-had-rep}
%\nonumber \\
\ea
and
\ba
{\rm Im} ~ I_\sigma^{(2n,0)}(q^2)_{\rm had} =  \pi
\delta(q^2-m_\pi^2){2n+1 \over 3}f_\pi^2 m_{0\pi}^\sigma m_{0\pi}^p \left<
\xi_\sigma^{2n}\right> +\pi {3 \over 8 \pi^2}{1\over
2n+3}q^2\theta(q^2-s_\pi^\sigma)~ .
\label{pi-sigma-had-rep}
%\nonumber \\
\ea
The correlation function in these two regions can be related by the dispersion relation,
\ba
{1\over\pi}\int ds {{\rm Im}~I(s)_{\rm had} \over s+Q^2} = I(-Q^2)_{\rm qcd}
\nonumber
\ea
In order to improve its convergence, we apply the
Borel transformation,
\ba
{1 \over \pi}{1 \over M^2}\int ds ~e^{-s/M^2}~{\rm Im}~I(s)_{\rm had} = \hat
L_M~I(-Q^2)_{\rm QCD} ~ ,
\label{borel-dispersion}
\ea
where $ ~M~ $ is Borel parameter.
Substituting (\ref{pi-p-qcd}) and (\ref{pi-p-had-rep}) into (\ref{borel-dispersion}) gives
the sum rules for the moments of $\phi_p^\pi$ :
\ba
\left< \xi_p^{2n}\right>
(m_{0\pi}^p)^2 &\makebox[0pt]{=}& { e^{m_\pi^2 /M^2}M^4 \over f_\pi^2}
\left\{{1 \over (2n +1)}{3 \over 8 \pi^2} \left[ 1-(1+{s_\pi^p
\over M^2})e^{-s_\pi^p/M^2}\right]\right.
\nonumber \\
&\makebox[0pt]{}&
+{1 \over 8}{\left< \dfrac{\alpha_s}{\pi}G^2 \right> \over M^4}
+{2n - 1 \over 2}{(m_u+m_d)\left< \bar q q\right> \over M^4}
\nonumber \\
&\makebox[0pt]{}& \left. +{16\pi \over 81}(16n^2+4n+21){\left<
\sqrt\alpha_s \bar q q\right>^2 \over M^6}\right\}~.
\label{pi-p}
\ea
Similarly, substituting Eqs.(\ref{pi-sigma-qcd}) and (\ref{pi-sigma-had-rep}) into Eq.(\ref{borel-dispersion})
gives the sum rules for the moments of $\phi_\sigma^\pi$ :
\ba
\left< \xi_\sigma^{2n}\right> m_{0\pi}^\sigma m_{0\pi}^p
&\makebox[0pt]{=}& 3 { e^{m_\pi^2 /M^2}M^4 \over f_\pi^2} \left\{
{1 \over (2n +1)(2n+3)}{3 \over 8 \pi^2} \left[ 1-(1+{s_\pi^\sigma
\over M^2})e^{-s_\pi^\sigma/M^2}\right] \right.
\nonumber \\
&\makebox[0pt]{}&
+~{1\over 24}~{1\over 2n+1}{\left< \dfrac{\alpha_s}{\pi}G^2\right> \over M^4}
+{1 \over 2}~{(m_u+m_d)\left< \bar q q\right> \over M^4}
\nonumber \\
&\makebox[0pt]{}& \left. +~{16\pi \over 81}~{16n^2+12n-7 \over
2n+1}~{\left< \sqrt\alpha_s \bar q q\right>^2 \over M^6} \right\}
\label{pi-sigma}~ ,
\ea
where $ ~s_\pi^p~ $ and $ ~s_\pi^\sigma~ $ are the threshold values to be chosen properly,
and the zeroth moment has been normalized to unit , i.e.,
$ \left< \xi_p^0 \right> = \left<\xi_\sigma^0 \right> = 1 $~.

\section{QCD sum rules for the moments of $ ~\phi_p^K~$ and $~\phi_\sigma^K~ $ of the kaon}

For the kaon, we should consider the difference
between $ s $ quark and $ u $ quark (i.e., the violation of the SU(3)
flavor symmetry). There is an asymmetry of the distribution of the longitudinal
momentum carried by $ s $ quark and $ u $ quark in the light cone framework. So
the odd moments of distribution amplitudes for $ K $ meson do not vanish.
The violation effects of the SU(3) flavor symmetry
for leading-twist distribution amplitudes of $K$ and/or $K^*$ meson
were considered in Ref.\cite{KMM}\,.
So in calculating the odd moments,
we retain all the corrections to order $ m_s^2 $\,.

Expanding equations (\ref{19}) and (\ref{20}) around $ z^2=0 $, one obtains
\ba
\left< 0 \left| \bar s(0)\gamma_5(i z\cdot \overleftrightarrow{D})^{n} u(0) \right| K^+(q) \right>
= -i f_K m_{0 K}^p \left< \zeta_p^{n} \right> (z\cdot q)^{n}
\label{kp-moment-add}
\ea
and
\ba
\left< 0 \left| \bar s(0)\sigma_{\mu \nu}\gamma_5 (iz\cdot
\overleftrightarrow{D})^{n+1} u(0) \right| K^+(q) \right> = -{n+1 \over
3}f_K m_{0 K}^\sigma \left< \zeta_\sigma^{n} \right> (q_\mu z_\nu
- q_\nu z_\mu) (z\cdot q)^{n}
\label{ksigma-moment-add}
\ea
respectively, and the moments are defined by :
\ba
\left< \zeta_p^n \right>=\frac{1}{2}\int_{-1}^{1}\zeta^n~\phi_p^K(\zeta)d\zeta~
,\quad\quad \left< \zeta_\sigma^n
\right>=\frac{1}{2}\int_{-1}^{1}\zeta^n~
\phi_\sigma^K(\zeta)d\zeta ~.
\ea
Similar to the pion case, the correlation functions for calculating the moments of the kaon are defined as,
\ba
(z\cdot q)^{n}~I_{K p}^{(n,0)}(q^2) \equiv -i\int d^4 x e^{iq\cdot x}
\left< 0 \left| {\rm T}\left \{ \bar s(x)\gamma_5(i z\cdot\overleftrightarrow{D})^{n} u(x),
   \bar u(0)\gamma_5 s(0)\right \}\right| 0 \right>
\label{23}
\ea
and
\ba
&& - i (q_\mu z_\nu - q_\nu z_\mu)
(z\cdot q)^n~I_{K \sigma}^{(n,0)}(q^2)
\nonumber \\
&&\quad \equiv -i\int d^4 x e^{iq\cdot x} \left< 0 \left| {\rm T}
\left \{ \bar s(x)\sigma_{\mu \nu}\gamma_5(iz\cdot
\overleftrightarrow{D})^{n+1} u(x), \bar u(0)\gamma_5 s(0) \right
\}\right| 0 \right> ~.
\label{24}
\ea
As discussed in the previous section, the correlation functions can be calculated
perturbatively in deep Euclidean region,
i.e., $Q^2=-q^2\gg 0$. Combined with
Eqs.(\ref{kp-moment-add}) and (\ref{ksigma-moment-add}),
we assume the hadronic spectrum representations of the above correlations as
follows,
\ba
{\rm Im} ~ I_{Kp}^{(n,0)}(q^2)_{\rm had} =
\pi \delta(q^2-m_K^2)f_K^2 (m_{0 K}^p)^2 \left< \zeta_p^{n}\right>
+\pi {3 \over 8 \pi^2}{1\over n+1}q^2\theta(q^2-s_K^p)
%\nonumber \\
\ea
and
\ba
{\rm Im} ~ I_{K\sigma}^{(n,0)}(q^2)_{\rm had} =  \pi
\delta(q^2-m_K^2){n+1 \over 3}f_K^2 m_{0 K}^\sigma m_{0 K}^p \left<
\zeta_\sigma^{n}\right> +\pi {3 \over 8 \pi^2}{1\over
n+3}q^2\theta(q^2-s_K^\sigma)~ .
%\nonumber \\
\ea
Employing the dispersion relation and Borel transformation as done in previous section,
the sum rules for the moments of $ ~\phi_p^K~ $ can be expressed in the following,
\ba
\left< \zeta_p^{2n}\right> (m_{0 K}^p)^2
&\makebox[0pt]{=}& { e^{m_K^2 /M^2}M^4 \over f_K^2}
\left\{{1\over (2n +1)}{3 \over 8 \pi^2} \left[ 1-(1+{s_K^p \over M^2})
e^{-s_K^p / M^2}\right]\right.
\nonumber \\
&\makebox[0pt]{}&
+{1 \over 8}{\left< \dfrac{\alpha_s}{\pi}G^2 \right> \over M^4}
+{\left[ (2n+1)m_s-2m_u \right]\left< \bar s s\right>
  + \left[ (2n+1)m_u-2m_s \right]\left< \bar u u\right>  \over 2 M^4}
\nonumber \\
&\makebox[0pt]{}& \left.
+{16\pi \over 81}(8n^2+2n-3){\alpha_s \left[ \left<  \bar s s \right>^2
                                      + \left<  \bar u u \right>^2 \right] \over M^6}
+ {16\pi \over 3}{ \alpha_s \left<  \bar s s \right>
                           \left<  \bar u u \right> \over M^6}
\right\}
\label{k-p-even}
\ea
and
\ba
\left< \zeta_p^{1}\right> (m_{0 K}^p)^2
&\makebox[0pt]{=}& { e^{m_K^2 /M^2}M^4 \over f_K^2}
\left\{ -{3\over 8 \pi^2}{m_s^2 \over M^2}\left( 1 - e^{-s_K^p / M^2}\right)
+{(m_s-m_u)\left[\left< \bar u u\right> + \left< \bar s s \right> \right]\over M^4}  \right.
\nonumber \\
&\makebox[0pt]{}&
\left.+{m_s^2 \left< \dfrac{\alpha_s}{\pi}G^2\right> \over 4 M^6}
+{4 \pi \over 27}{m_s^2\over M^2}
 { 36~\alpha_s\left<  \bar u u \right>\left<  \bar s s \right> - 4~\alpha_s\left<  \bar u u \right>^2
   \over M^6}\right\}~.
\label{k-p-odd}
\ea
For the twist-3 amplitude $ ~\phi_\sigma^K~ $, we make a similar calculation
according to the above procedure and the sum rules for the moments of
$ ~\phi_\sigma^K~ $ become
\ba
\left< \zeta_\sigma^{2n}\right> m_{0 K}^p m_{0 K}^\sigma
&\makebox[0pt]{=}& 3 { e^{m_K^2 /M^2}M^4 \over f_K^2} \left\{{1
\over (2n +1)(2n+3)}{3 \over 8 \pi^2} \left[ 1-(1+{s_K^\sigma
\over M^2})e^{-s_K^\sigma/M^2}\right]\right.
\nonumber \\
&\makebox[0pt]{}&
+{1 \over 24(2n+1)}{\left< \dfrac{\alpha_s}{\pi}G^2 \right> \over M^4}
+{m_s\left< \bar s s\right> + m_u\left< \bar u u\right>  \over 2 M^4}
\nonumber \\
&\makebox[0pt]{}& \left.
+{16\pi \over 81}(4n+1){\alpha_s \left[ \left<  \bar s s \right>^2
                                      + \left<  \bar u u \right>^2 \right] \over M^6}
- {16\pi \over 9(2n+1)}{ \alpha_s \left<  \bar s s \right>
                                  \left<  \bar u u \right> \over M^6}
\right\}
\nonumber \\
\label{k-sigma-even}
\ea
and
\ba \left< \zeta_\sigma^{1}\right>
m_{0 K}^p m_{0 K}^\sigma &\makebox[0pt]{=}& {3\over 2}{ e^{m_K^2
/M^2}M^4 \over f_K^2} \left\{ -{ 1 \over 4 \pi^2}{m_s^2 \over
M^2}\left( 1 - e^{-s_K^\sigma/M^2}\right) +{ m_s\left< \bar s s
\right> - m_u \left< \bar u u\right>  \over M^4}  \right.
\nonumber \\
&\makebox[0pt]{}&
+{m_s^2 \left< \dfrac{\alpha_s}{\pi}G^2\right> \over 6 M^6}
(\ln{M^2\over \mu^2}+1-\gamma_E)
- {1\over 3} {m_s g_s\left< \bar s \sigma G s \right> \over M^6 }
\nonumber \\
&\makebox[0pt]{}&
\left. + {32\pi \over 27}{\alpha_s \left[ \left< \bar s s \right>^2 - \left< \bar u u \right>^2 \right]\over M^6}
\right\}~,
\label{k-sigma-odd}
\ea
where $ ~\gamma_E=0.577~216\cdots~ $ is the Euler constant, $ ~s_K^p~ $ and $ ~s_K^\sigma~ $ in the
above equations are the threshold values to be chosen properly, and the zeroth moments has been
normalized, $ \left< \zeta_p^0 \right> = \left<\zeta_\sigma^0 \right> = 1 $~.

\section{Numerical analysis}

To analyse the sum rules (\ref{pi-p}),(\ref{pi-sigma}),(\ref{k-p-even})-(\ref{k-sigma-odd})
numerically, we take the input parameters as usual :
$ f_K = 0.160{\rm ~GeV}$,
$f_\pi = 0.133{\rm ~GeV},~ m_s=0.156{\rm ~GeV},~ m_u=0.005{\rm ~GeV},$
$~m_d=0.008{\rm ~GeV},~\left< \bar u u \right> = \left< \bar d d \right> = -(0.24{\rm ~GeV})^3,$
$~\left< \bar s s \right> = 0.8~ \left< \bar u u \right>,$
$~g_s \left< \bar s \sigma G s \right> = -0.00885~{\rm GeV}^5,
~\left< {\displaystyle{\alpha_s \over \pi}} GG \right> = 0.012{\rm ~GeV}^4,
~\alpha_s(1{\rm GeV})=0.5~.$ The renormalization scale $ \mu = M $ is assumed in the following analysis.

As to the threshold values $ s_{\pi,K}^{p,\sigma} $ in the sum
rules, they can be taken to the mass square of the first exciting
states in the corresponding channel. Although the windows
become broader when the $ s_{\pi, K}^{p,\sigma} $ are larger,
the threshold values can not exceed the first exciting states.
In order to get maximum stability of the sum rules, they are taken as
the mass square of the first exciting states, i.e.,
$$
s_\pi^{p,\sigma}=(1.3~{\rm GeV})^2,\quad\quad\quad
s_K^{p,\sigma}=(1.46~{\rm GeV})^2
$$
where the first exciting state is $ \pi'(1300) $ for the pion case, and that for
kaon case is $ K(1460) $ \cite{k-1460}.

\subsection{Determination of the normalization constants}
\begin{figure}[ht]
\setlength{\unitlength}{1cm}
\begin{minipage}[t]{7.5cm}
\begin{picture}(7.5,4.5)
\includegraphics*[scale=0.85,angle=0.]{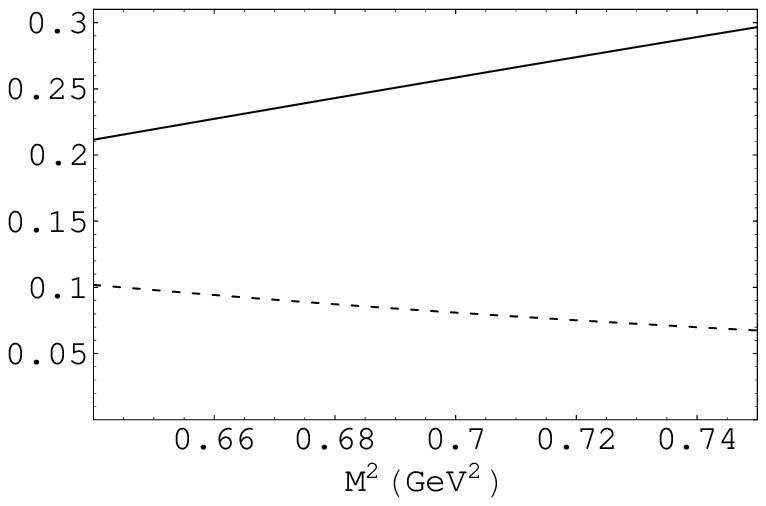}
\end{picture}\par
(a)
\end{minipage}
\hfill
\begin{minipage}[t]{7cm}
\begin{picture}(7,4.5)
\includegraphics*[scale=0.85,angle=0.]{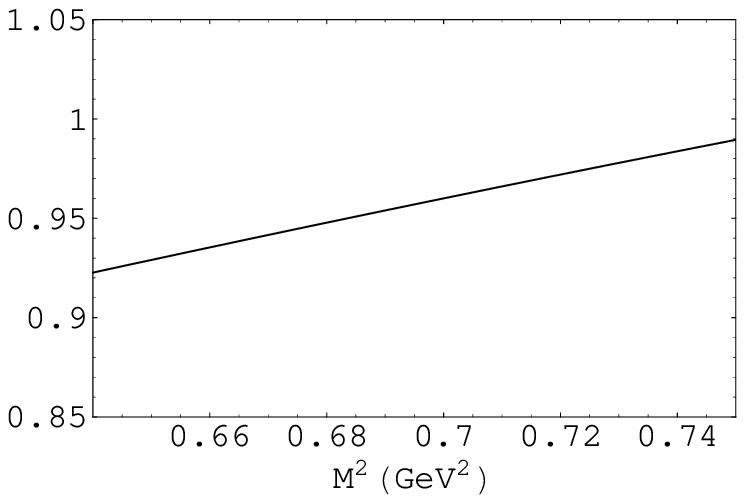}
\end{picture}\par
(b)
\end{minipage}
\caption{\label{fig-m0-p}(a) The window for the normalization constant $ m_{0\pi}^p $
without $\alpha_s$ correction in the perturbative part in the sum rule.
The dashed line is the ratio of the dimension-six condensate contribution
to the total sum rule($n=0$) and the solid line is the ratio of the continuum contribution
to the total sum rule($n=0$) . (b) The corresponding values
of $ m_{0\pi}^p $ within the window.}
\end{figure}

For each distribution amplitude, we
introduce a corresponding parameter (e.g., $ m_{0\pi}^p $ for $ \phi_p^\pi $).
These parameters are
normalization constants which normalize the zeroth moments to one.
Their values can be determined from the sum rules
(\ref{pi-p}), (\ref{pi-sigma}), (\ref{k-p-even}) and (\ref{k-sigma-even})
with $n=0$.
Take $ m_{0\pi}^p $ as an
example. To identify a Borel window ($ M^2 $) for the
sum rule of $ m_{0\pi}^p $, one requires
that the continuum contribution is less than 30\% and the
dimension-six condensate contribution is less than 10\%.
This requirement leads to a window $ M^2 \in(0.64~,~0.75)~{\rm GeV}^2 $
and one can find $ m_{0\pi}^p = 0.96\pm 0.03~{\rm GeV} $ within this window.
And the results are plotted in the Fig \ref{fig-m0-p}(a) and Fig \ref{fig-m0-p}(b)
for $ m_{0\pi}^p $.

%%%%%%%%%%%%%%%%%     figures     %%%%%%%%%%%%%%%%%%%%
%%%%%%%%%%%%%%%%%     figures     %%%%%%%%%%%%%%%%%%%%%%

The same procedure can be applied to get other parameters
$ m_{0\pi}^\sigma,~m_{0 K}^{p,\sigma} $, the numerical results are listed in Table \ref{table-1} .
The continuum contribution to the sum rules are required to be less than 30\%
and the dimension-six contribution is required to be less than 16\% for $m_{0\pi}^\sigma$,
10\% for $m_{0 K}^p$ and $m_{0 K}^{\sigma}$\,.
It should be pointed out that when the $\alpha_s$ correction to the perturbative part of the sum
rule for $ m_{0\pi}^p,~m_{0 K}^p $ are taken into account \cite{back-ground-field-2},
their values will be increased by 15-20$\%$\,. For example,
$ m_{0\pi}^p= 1.10\pm 0.08~{\rm GeV} $ and $ m_{0 K}^p= 1.25\pm 0.15~{\rm GeV}$.

One can see from above that $m_{0\pi}^\sigma$ is smaller than $m_{0\pi}^p$ about 30\%. The
main reason is that the opposite sign of the dimension-six condensate terms in
Eqs. (\ref{pi-p}) and (\ref{pi-sigma}).
For the kaon case, the approximate 30\% difference between $m_{0 K}^{\sigma}$
and $m_{0 K}^p $ is due to the same reason(see Eqs.(\ref{k-p-even})
and (\ref{k-sigma-even})).

%%%%%%%%%%%%%%%%%     tables      %%%%%%%%%%%%%%%%%%%%%%

\begin{table}
\bc
\begin{tabular}{|c|c|c|c|c|}
\hline
  & $ \phi_p^\pi $ & $ \phi_\sigma^\pi $ & $ \phi_p^K $ & $ \phi_\sigma^K $ \\
\hline
 $m_0~({\rm GeV})$ & $ 0.96\pm 0.03 $ & $ 0.67\pm 0.06 $
                  & $ 1.06\pm 0.09 $ & $ 0.71\pm 0.09 $ \\
\hline
 $M^2$~$({\rm GeV^2})$  & 0.64-0.75 & 0.60-0.68 & 0.58-0.93 & 0.66-0.83 \\
\hline
\end{tabular}
\ec
\caption{\label{table-1}Normalization constants $m_0$s and the corresponding Borel windows for
the distribution amplitudes $ \phi_{p,\sigma}^{\pi}$ and $ \phi_{p,\sigma}^K$
without $\alpha_s$ correction in the perturbative parts in the sum rules. }
\end{table}

%%%%%%%%%%%%%%%%%     tables      %%%%%%%%%%%%%%%%%%%

It was shown that the normalization constants
for the twist-3 distribution amplitudes can be
obtained from equations of motion \cite{light-cone-non-leading-twist-def-1}.
So at this point, we would like to compare our
results with those obtained by equations of motion and
judge upon the accuracy of the sum rules presented above.
From the equations of motion follows the normalization constant
$m_{0\pi}^p \to  \mu_\pi \triangleq m_\pi^2/\bar m$ for $ \phi_p^{\pi}$ and
$m_{0\pi}^\sigma \to \widetilde\mu_\pi\triangleq\mu_\pi-\bar m $
for $ \phi_\sigma^{\pi}$\,, where
$\bar m= m_u +m_d$. Furthermore, the $Q^2$ dependence of the quark
mass $\bar m$ can be written as:
$\bar m(Q^2)=[\ln (\mu^2/\Lambda^2)/\ln(Q^2/\Lambda^2)]^{4/9}\,\bar m(\mu^2)$
which can be obtained from the anomalous dimension of quark mass.
Thus we find $\mu_\pi(1\,\hbox{GeV}^2)\approx 1.48\hbox{\,GeV}$ and
$\widetilde\mu_\pi(1\,\hbox{GeV}^2)\approx 1.47\hbox{\,GeV}$
as we take $\bar m(4 \,\hbox{GeV}^2)=11 \,\hbox{MeV}$\,.
For the kaon case, we take $(m_u+m_s)(4 \,\hbox{GeV}^2)=140 \,\hbox{MeV}$\,.
From the equation of motion we have
$\mu_K(1\,\hbox{GeV}^2)=m_K^2/(m_u+m_s)(1\,\hbox{GeV}^2)
\approx 1.45\hbox{\,GeV}$ and
$(\mu_K-(m_s+m_u))(1\,\hbox{GeV}^2)\approx 1.28\hbox{\,GeV}$\,.
In the above statement, QCD scale $\Lambda=250\,\hbox{MeV}$ is assumed
and $n_f=3$ flavors are taken into account.
One can see that the deviation of $m_{0\pi}^p$ from
$\mu_\pi$ is aboat 26\% and the deviation
of $ m_{0 K}^p$ from $\mu_K$ is less than 15\%
($\alpha_s$ corrections to the perturbative parts are included).

If the $\alpha_s$ correction to the perturbative part is 15-20\% and
these corrections make the normalization constants increasing,
one expect that the deviation of $ m_{0 \pi}^\sigma$
from $\widetilde\mu_\pi$ is about 45\% and the deviation
of $ m_{0 K}^\sigma$ from $\mu_K-(m_u+m_s)$ is about 33\%\,.

\subsection{Determination of the second moment of $ \phi_{p,\sigma}^\pi $ and
the odd moment of $ \phi_{p,\sigma}^K $ }

\begin{figure}[ht]
\setlength{\unitlength}{1cm}
\begin{minipage}[t]{7cm}
\begin{picture}(7,4.5)
\includegraphics*[scale=0.85,angle=0.]{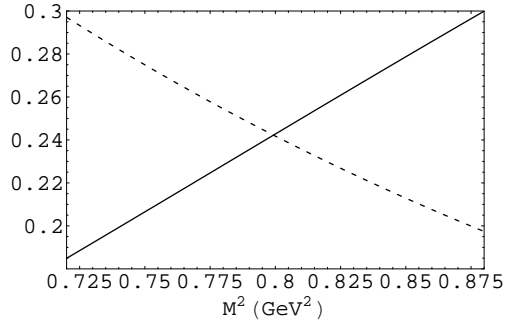}
\end{picture}\par
(a)
\end{minipage}
\hfill
\begin{minipage}[t]{7cm}
\begin{picture}(7,4.5)
\includegraphics*[scale=0.85,angle=0.]{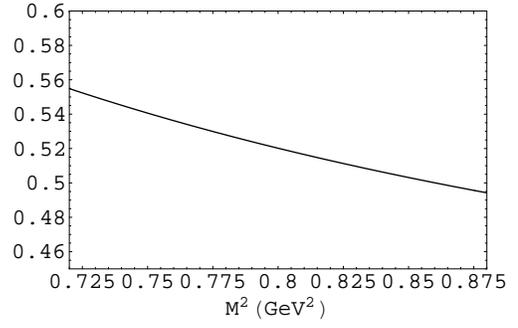}
\end{picture}\par
(b)
\end{minipage}
\caption{\label{fig-pi-p-second}
(a) the window for the moment $ \langle \xi_p^2 \rangle $, the dashed and the solid line indicate the ratio of
the contribution of dimension-6 condensates and  continuous states in the total sum rule respectively;
(b) the moment $ \langle \xi_p^2 \rangle $ within the Borel window. }
\end{figure}

Let us consider the second moments of $ \phi_p^\pi $ and $ \phi_\sigma^\pi $ for the pion.
Just as the determination of normalization constants in the above paragraphs,
one should find a window for each moment in the corresponding sum rule.
The Borel windows in Table \ref{table-2} are obtained under requirement that
both the contributions from continuous states and the dimension-six
condensate are less than 30$\%$\,.
 As an example, we plot
the results for the moment $ \left< \xi_p^2 \right> $ in the Fig \ref{fig-pi-p-second}(a) and
the Fig \ref{fig-pi-p-second}(b) and the numerical results are
listed in Table \ref{table-2}.

%%%%%%%%%%%%%%%%%%%%%%%%%%%%%%%%%%%%%%%%%%%%%%     figures     %%%%%%%%%%%%%%%%%%%%%%%%%%%%%%%%%%%%%%%
%%%%%%%%%%%%%%%%%%%%%%%%%%%%%%%%%%%%%%%%%%%%%%     figures     %%%%%%%%%%%%%%%%%%%%%%%%%%%%%%%%%%%%%%%

\begin{figure}[th]
\setlength{\unitlength}{1cm}
\begin{minipage}[t]{7cm}
\begin{picture}(7,4.5)
\includegraphics*[scale=0.85,angle=0.]{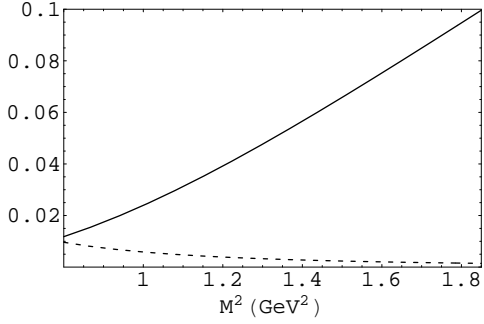}
\end{picture}\par
(a) $\langle \zeta_p^1 \rangle$
\end{minipage}
\hfill
\begin{minipage}[t]{7cm}
\begin{picture}(6.5,4.5)
\includegraphics*[scale=0.85,angle=0.]{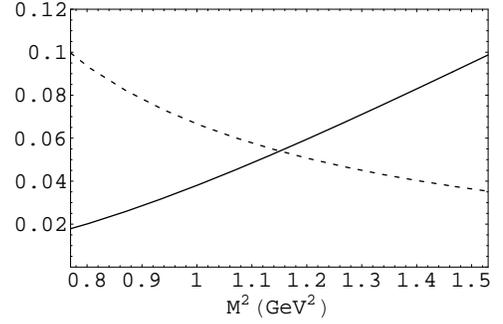}
\end{picture}\par
(b) $\langle\zeta_\sigma^1\rangle$
\end{minipage}
\caption{\label{fig-k-p-sigma}
The windows for $\langle \zeta_p^1 \rangle$ and  $\langle\zeta_\sigma^1\rangle$.
The dashed  and the solid lines indicate the ratios of the contribution of dimension-6 condensates and
the continuous states in the corresponding total sum rule respectively. }
\end{figure}

Now we turn to the determination of the first moments of $\phi_{p,\sigma}^K$.  The
contribution from the dimension-6 condensate and the continuous states of
$\langle \zeta_p^1 \rangle$ and $\langle\zeta_\sigma^1\rangle$ are plotted in
Fig.\ref{fig-k-p-sigma}.
For $ \left< \zeta_p^1 \right> $\,, the dimension-six contribution
is less than 1\% and the continuum contribution is less than 10\%\,.
For $ \left< \zeta_\sigma^1 \right> $\,, the contribution of dimension-six condensate
and continuous states are less than 10\%\,.

With these windows we can get the values of the corresponding moments. These results are
listed in Table \ref{table-2}.

%%%%%%%%%%%%%%%%%     figures     %%%%%%%%%%%%%%%%%%%%%%%
%%%%%%%%%%%%%%%%%     figures     %%%%%%%%%%%%%%%%%%%%%%%

%%%%%%%%%%%%%%%%%     tables      %%%%%%%%%%%%%%%%%%

\begin{table}
\bc
\begin{tabular}{|c|c|c|c|c|}
\hline
  & $ \left<\xi_p^2\right> $ & $ \left<\xi_\sigma^2\right> $
  & $ \left<\zeta_p^1\right> $ & $ \left<\zeta_\sigma^1\right> $ \\
\hline
  & $ 0.52\pm 0.03  $ & $ 0.34\pm 0.03 $
  & $ -0.10\pm 0.03 $ & $ -0.13\pm 0.04 $ \\
\hline
 $M^2$~$({\rm GeV^2})$  & 0.72-0.88 & 0.71-0.84 & 0.80-1.85  & 0.77-1.53 \\
\hline
\end{tabular}
\ec
\caption{\label{table-2} Second moments of $ \phi_{p,\sigma}^\pi $, odd moments of
$ \phi_{p,\sigma}^K $ and their corresponding Borel windows.
}
\end{table}

%%%%%%%%%%%%%%%%     tables      %%%%%%%%%%%%%%%%%%%%%%%

\subsection{Determination of the fourth moment of $ \phi_{p,\sigma}^\pi $ and the second moment of
$ \phi_{p,\sigma}^K $}

\begin{figure}[th]
\setlength{\unitlength}{1cm}
\begin{minipage}[t]{7cm}
\begin{picture}(7,4.5)
\includegraphics*[scale=0.85,angle=0.]{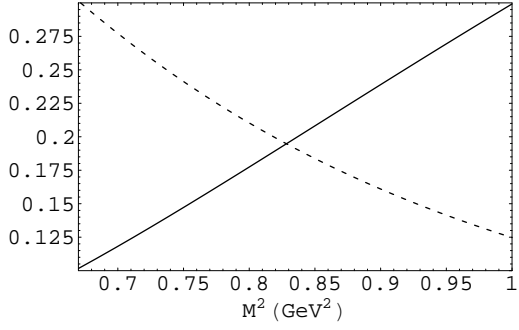}
\end{picture}\par
(a) $\langle \zeta_p^2 \rangle$
\end{minipage}
\hfill
\begin{minipage}[t]{7cm}
\begin{picture}(6.5,4.5)
\includegraphics*[scale=0.85,angle=0.]{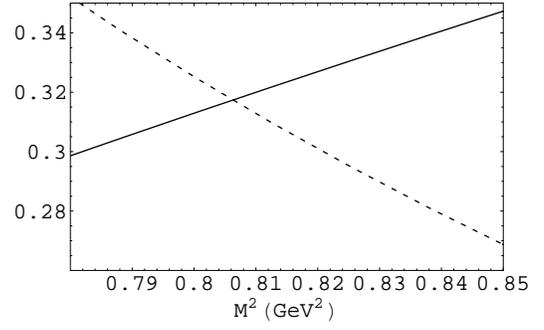}
\end{picture}\par
(b) $\langle\zeta_\sigma^2\rangle$
\end{minipage}
\vskip 0.4cm
\begin{minipage}[t]{7cm}
\begin{picture}(7,4.5)
\includegraphics*[scale=0.85,angle=0.]{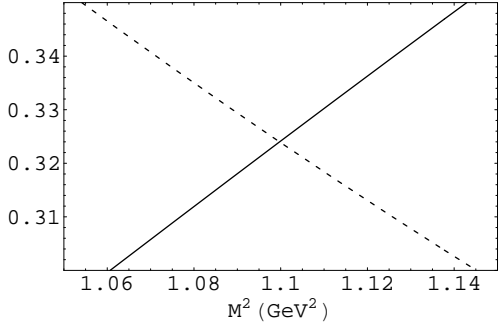}
\end{picture}\par
(c) $\langle \xi_p^4 \rangle$
\end{minipage}
\hfill
\begin{minipage}[t]{7cm}
\begin{picture}(6.5,4.5)
\includegraphics*[scale=0.85,angle=0.]{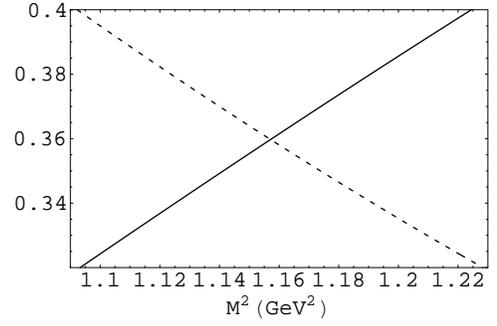}
\end{picture}\par
(d) $\langle\xi_\sigma^4\rangle$
\end{minipage}
\caption{\label{last-four-moments-window}
The windows for the second moments of $ \phi_p^K , \phi_\sigma^K $ and the fourth
moments of $ \phi_p^\pi  , \phi_\sigma^\pi $.
The dashed  and the solid lines indicate the ratios of the contribution of dimension-6 condensates and
the continuous states in the corresponding total sum rule respectively. }
\end{figure}

Now we consider the second moment $ \left< \zeta_p^2 \right> $
of $ \phi_p^K $ for the $ K $ meson.
The Borel window for $ \left< \zeta_p^2 \right> $
is shown in the Fig.\ref{last-four-moments-window}(a)
when the contribution of continuous states and the dimension-six condensate
are less than 30\%\,.
The numerical results are listed in Table \ref{table-3}.

However, for the fourth ($ n=2 $) moments of
$ \phi_p^\pi  , \phi_\sigma^\pi $ of the $ \pi $ meson,
we can not find the Borel windows when
the contribution of continuous states and the dimension-six condensate
are required to be less than 30\%\,.
For the second ($ n=1 $) moment of $ \phi_\sigma^K $ of the $ K $ meson,
we find that the Borel window is very narrow when
the contribution of continuous states and the dimension-six condensate
are required to be less than 30\%\,.
As we relax the requirement that the contribution of continuous states
and the dimension-six condensate are less than 35\%, the Borel windows
for $\langle \zeta_\sigma^2 \rangle$ and $\langle \xi_p^4 \rangle$
can be found.
For $\langle \xi_\sigma^4 \rangle$\,, one can find the Borel window
only when the contribution of continuous states and the dimension-six condensate
are less than 40\%\,.
The above windows are shown in
Fig. \ref{last-four-moments-window}(b)-\ref{last-four-moments-window}(d)\,.

The values of these moments within their corresponding windows are listed in Table \ref{table-3}.

%%%%%%%%%%%%%%%%%%%%     figures     %%%%%%%%%%%%%%%%%%%%%%%%%%
%%%%%%%%%%%%%%%%%     figures     %%%%%%%%%%%%%%%%%%%%%%%

%%%%%%%%%%%%%%%%%%%%%%%%    tables      %%%%%%%%%%%%%%%%%%%%
\begin{table}
\bc
\begin{tabular}{|c|c|c|c|c|}
\hline
  & $ \left<\xi_p^4\right> $ & $ \left<\xi_\sigma^4\right> $
  & $ \left<\zeta_p^2\right> $ & $ \left<\zeta_\sigma^2\right> $ \\
\hline
  & $ 0.44\pm 0.01 $ & $ 0.20\pm 0.01 $  & $ 0.43\pm 0.04 $ & $ 0.173\pm 0.002 $ \\
\hline
$M^2$~$({\rm GeV^2})$  & 1.06-1.14 & 1.08-1.22 & 0.67-1.00 & 0.78-0.85  \\
\hline
\end{tabular}
\ec
\caption{\label{table-3}Fourth moments of $ \phi_{p,\sigma}^\pi $ and second moments of
$ \phi_{p,\sigma}^K $. But note that the values of $ \langle \xi_p^4 \rangle $,
$ \langle\zeta_\sigma^2 \rangle$ given in this table is under the requirement of 35$\%$ uncertainty
and $ \langle \xi_\sigma^4\rangle $ is under the requirement of 40$\%$ uncertainty. }
\end{table}

%%%%%%%%%%%%%%%%%%%%%%     tables      %%%%%%%%%%%%%%%%%%

\subsection{From two-particle distribution amplitudes to three-particle distribution amplitude}

There are three twist-3 distribution amplitudes $ \phi_p^\pi $\,, $ \phi_\sigma^\pi $
and $ \phi_{3\pi} $ for $\pi$ meson. As shown in
Ref.\cite{light-cone-non-leading-twist-def-1}\,, they are not independent.
By employing equations of motion in QCD, one can obtain some relations between
them. For the pion, the relations between the twist-3 distribution amplitudes
of the lowest fork state
and the 3-particle one are given in Ref.\cite{light-cone-non-leading-twist-def-1}\,.
They obtained two distribution amplitudes of the lowest fork state $ \phi_{p,\sigma}^\pi $
from the 3-particle distribution amplitude $\phi_{3\pi}$ which was given by a
direct calculation in the QCD sum rule method\cite{3-particle}\,.
Contrarily, we use the relations from equations of motion to see what  we can
say about the 3-particle distribution amplitudes with the above results of the
distribution amplitudes  of the lowest fork state as input. The results can also be
compared with those obtained
by QCD sum rule directly\cite{3-particle}\,.
The cross checks in these calculations are helpful to judge upon the accuracy
of the sum rules.

First, let's discuss the pion case.
The three-particle distribution amplitude of the $\pi$ meson can
be defined as,
\ba
&&\langle 0 | \bar d(x) \sigma_{\mu\nu}\gamma_5
    g G_{\alpha\beta}(-vx)  u(-x)| \pi^+(q)\rangle
\nonumber\\&&\qquad
= i f_{3\pi}\big[q_\alpha(q_\mu\delta_{\nu\beta}-q_\nu\delta_{\mu\beta})
               - (\alpha\leftrightarrow \beta)\big]
\int \mathscr{D}\alpha_i\, e^{i qx(-\alpha_1+\alpha_2+v\alpha_3)}\phi_{3\pi}(\alpha_i)
\ea
where $\mathscr{D}\alpha_i
=d\alpha_1d\alpha_2d\alpha_3\delta(\alpha_1+\alpha_2+\alpha_3-1)$\,.
There are the system of  recurrence relations for the moments
$\langle \xi_p^n\rangle$ and $\langle \xi_\sigma^n\rangle$
\cite{light-cone-non-leading-twist-def-1}\,:
\ba
\langle \xi_p^n\rangle &=& \delta_{n 0}
+ {n-1\over n+1} \langle \xi_p^{n-2}\rangle
+2 R_p(n-1)
\int_{-1}^1 dv \langle\!\langle(\alpha_2 -\alpha_1+v\alpha_3)^{n-2}\rangle\!\rangle
\nonumber\\&&
- 2R_p{(n-1)(n-2)\over n+1}
\int_{-1}^1 dv\,v \langle\!\langle(\alpha_2 -\alpha_1+v\alpha_3)^{n-3}\rangle\!\rangle
\\
\langle \xi_\sigma^n\rangle &=& \delta_{n 0}
+ {n-1\over n+3} \langle \xi_\sigma^{n-2}\rangle
+6 R_\sigma{n-1\over n+3}
\int_{-1}^1 dv \langle\!\langle(\alpha_2 -\alpha_1+v\alpha_3)^{n-2}\rangle\!\rangle
\nonumber\\&&
- 6 R_\sigma{n\over n+3}
\int_{-1}^1 dv\,v \langle\!\langle(\alpha_2 -\alpha_1+v\alpha_3)^{n-1}\rangle\!\rangle
\label{recurrence}
\ea
where $\langle\!\langle(\alpha_2 -\alpha_1+v\alpha_3)^n\rangle\!\rangle
=\int \mathscr{D}\alpha^{}_i\, \phi^{}_{3 \pi}(\alpha^{}_i)\,
(\alpha_2 -\alpha_1+v\alpha_3)^n$ defines the moments of 3-particle
distribution amplitude.
Instead of taking $R_p = R_\sigma =R$ as in Ref.\cite{light-cone-non-leading-twist-def-1}\,,
we introduce them seperately,
\ba
R_p={1\over m_{0\pi}^p}{f_{3\pi}\over  f_\pi}\qquad \hbox{and}\qquad
R_\sigma={1\over m_{0\pi}^\sigma }{f_{3\pi}\over f_\pi}
\nonumber
\ea
Now, we take second moments into account. The above relation can be reduced
to
\ba
\langle \xi_\sigma^2\rangle &=& {1\over 5} \langle \xi_\sigma^0\rangle
+{12\over 5}R_\sigma - {8\over 5} R_\sigma \langle\!\langle\alpha_3 \rangle\!\rangle
\nonumber
\ea
and
\ba
\langle \xi_p^2\rangle &=& {1\over 3} \langle \xi_p^0\rangle + 4 R_p\,.
\nonumber
\ea
which gives, from Table \ref{table-1} and Table \ref{table-2},
\ba
\langle\!\langle\alpha_3 \rangle\!\rangle = (0.13\,, \,0.27)
\qquad \hbox{and}\qquad
f_{3\pi}=(0.0049\,,\, 0.0067)\hbox{GeV}^2\,.
\ea
At this point, we compare the moment $\langle\!\langle\alpha_3 \rangle\!\rangle$
and the $f_{3\pi}$ with those calculated directly by the sum
rule method in Ref. \cite{3-particle}\,:
$\langle\!\langle\alpha_3 \rangle\!\rangle = (0.06\,, \,0.22)$\,,
$f_{3\pi}\approx 0.0035\, \hbox{GeV}^2$\,.
One can see that the results from the two approaches are compatible to the
order of magnitude.

From the analysis in previous section, we have shown that the fourth
moments $\langle \xi_p^4\rangle$ and $\langle \xi_\sigma^4\rangle$
can not be obtained in a reliable way, so we do not use them to
give the other moments, i.e.,
$\langle\!\langle\alpha_1^2 \rangle\!\rangle$ and
$\langle\!\langle\alpha_1\alpha_2 \rangle\!\rangle$\,, etc..

Now we turn to the $K$ meson case. Similar to the pionic case, one
can define a three-particle distribution amplitude
\ba
&&\langle 0 | \bar s(x) \sigma_{\mu\nu}\gamma_5
    g G_{\alpha\beta}(-vx)  u(-x)| K^+(q)\rangle
\nonumber\\
&&\quad = i f_{3K}[q_\alpha(q_\mu\delta_{\nu\beta}-q_\nu\delta_{\mu\beta})
               - (\alpha\leftrightarrow \beta)]
\int \mathscr{D}\alpha_i\, e^{i qx(-\alpha_1+\alpha_2+v\alpha_3)}\phi_{3K}(\alpha_i)\,.
\ea
Following Ref.\cite{light-cone-non-leading-twist-def-1}, a similar recurrence relation
can be obtained. As the first and second moment
of $\phi_{p,\sigma}^K$ are taken into account, the recurrence
relation can be truncated to three equations,
\ba
\left<\zeta_\sigma^1\right>
&=& {3\over 4} {R_\sigma'\over R_p'}\left<\zeta_p^1\right>
\label{check}
\\
\left<\zeta_\sigma^2\right>
 &=& {3\over 5} {R_p'\over R_\sigma'}\left<\zeta_p^2\right>
   - {8\over 15} R_p' \langle\!\langle\alpha_3 \rangle\!\rangle_K
\label{k-eq-1}
\\
\left<\zeta_p^2\right>
&=& {1\over 3} {R_p'\over R_\sigma'}\left<\zeta_\sigma^0\right>
+4 R_p'
\label{k-eq-2}
\ea
where $R_{p,\sigma}'=f_{3K}/(f_K m_{0K}^{p,\sigma})$\,,
and the primes on the $R$s and the subscript $K$ indicate that  the quantities are
related to $K$ meson.
From Table \ref{table-1}, we have ${R_\sigma'/ R_p'}\approx 1.06/0.71$\,, so
the equation (\ref{check}) is a direct constraint of the two fist moments.
Our calculation (see Table \ref{table-2}) shows that the left hand side of Eq.(\ref{check})
is about $-0.13$ and the right hand side is about $-0.11$. It can be seen that
this equation is approximately fufilled.
Solving the last two equations (\ref{k-eq-1}) and (\ref{k-eq-2}), we can
obtain $f_{3K}$ and $\langle\!\langle\alpha_3 \rangle\!\rangle_K$:
\ba
f_{3K}=(0.0071\,,\, 0.0105)\,\hbox{GeV}^2\,,\qquad
\langle\!\langle\alpha_3 \rangle\!\rangle_K=(5.01\,,\, 5.37)
\ea
To determine more moments of the three-particle distribution amplitude, we have to
include higher moments of the two-particle distribution amplitudes.
However, one can not guarantee the convergence of the operator expansion for
bigger $n$\,.

\section{Summary and discussion}

In this paper we calculate the first three moments of the twist-3
distribution amplitudes $\phi_{p,\sigma}^\pi$ and $\phi_{p,\sigma}^K$ by using the QCD sum rules.
It has been shown that the first three moments of $ \phi_p^K $ and the first two
moments of $ \phi_p^\pi $ and $ \phi_\sigma^{\pi,K} $ of the pion and kaon can be
obtained with 30$\%$ uncertainty.
The fourth moments $\left<\xi_{p,\sigma}^4\right>$
of $\phi_{p,\sigma}^\pi$ and the second moment $\left<\zeta_\sigma^2\right>$
of $\phi_\sigma^K$ can be obtained under 35\%-40$\%$ uncertainty.
When the $\alpha_s$ corrections (we take them from Ref.\cite{back-ground-field-2})
to the perturbative part of $m_{0 K}^p~,~ m_{0\pi}^p$ are included,
we find that the values of $m_{0\pi}^p$ and $m_{0 K}^p$ are increased
(and the corresponding Borel windows becomes a little narrower)
to $ m_{0 K}^p= 1.25\pm 0.15~{\rm GeV}$ and $ m_{0\pi}^p= 1.10\pm 0.08~{\rm GeV} $.
It may be expected that the $\alpha_s$ corrections to the perturbative parts
in the sum rules for $m_{0 K}^{\sigma}$ and $m_{0\pi}^{\sigma}$
will be about 15-20$\%$.
%Consequently, the parameters $m_{0 K}^{p,\sigma}$ and $m_{0\pi}^{p,\sigma}$
%are smaller than those required by the equations of motion.

As to the normalization constants $m_{0\pi}^{p,\sigma}$ and $m_{0K}^{p,\sigma}$\,,
our calculated results show that they are smaller than the values which are given
by the equations of motion and at the same time, the calculated $m_{0\pi, K}^{\sigma}$ are smaller than
the corresponding $m_{0\pi, K}^{p}$\,.
These deviation can be traced to the non-perturbative condensate effects
(see the sum rules
(\ref{pi-p}), (\ref{pi-sigma}), (\ref{k-p-even}) and (\ref{k-sigma-even})
for the normalization constants), in particular,
the dimension-six condensate terms in opposite sign lead to
about 30\% difference between these normalization constants.
On the other hand, from the sum rules, one can see that the contribution from
the continuous state grow too fast which prevent us from taking $M^2$
to be larger values(larger $M^2$ will lead to bigger values of the normalization constants),
and then the window for the sigma-sum-rules($m_{0\pi, K}^{\sigma}$)
are much narrower than the non-sigma sum rules($m_{0\pi, K}^p$).
So we think the smaller values of $m_{0\pi, K}^{\sigma}$ may be
related to our approximation in the hadronic spectrum representation.

Furthermore, we calculate the moments of the quark-antiquark-gluon distribution
amplitude from the numerical results on the distribution amplitudes of the lowest Fork
state by applying the exact equations of motion and compare our results with those
from Ref.\cite{3-particle}. The comparison shows that they are compatible with each other
to the order of magnitude.  It is helpful to improve the accuracy of the QCD sum
rules approach for getting more precise information on the twist-3 distribution
amplitudes.

These moments can provide several constraints upon the twist-3 distribution amplitudes.
These constraints will be helpful for building the model of the distribution amplitude. For
example, Ref \cite{anomalous-twist-3} suggests a model for the twist-3 wave function of
the pion based on the QCD sum rule calculation to get a more realistic contribution to the
pion form factor. Here we discuss the distribution amplitude $\phi_p^K$ of the
kaon(since the first three moments can be obtained reliablely).
As usual, we expand the distribution amplitudes in
Gegenbauer's polynomials and use the moments to determine their first few coefficients
in a truncated form : $\phi_p^K(\zeta)=\sum_{n=0}^2C^{1/2}_n(\zeta)a_n$.
From the three moments of $\phi_p^K$ (see Table \ref{table-2} and Table \ref{table-3}), we have
the twist-3 distribution amplitude approximately,
\ba
\label{eq-phi-k-p-DA}
\phi_p^K(\zeta)=1 - 0.30~ C_1^{1/2}(\zeta) + 0.73~ C_2^{1/2}(\zeta)
\ea
which is asymmetric since the $SU(3)_f$
symmetry broken are taken into account.

\begin{acknowledgments}
We would like to thank A. Khodjamirian for his helpful communication.
This work was supported in part by the National Science Foundation of China, 10275070.
\end{acknowledgments}

%\newpage

\end{document}